# PHASE–TRANSITION THEORY OF INSTABILITIES.
# I. SECOND–HARMONIC INSTABILITY AND BIFURCATION POINTS

Dimitris M. Christodoulou[1], Demosthenes Kazanas[2], Isaac Shlosman[3,4], and Joel E. Tohline[5]

## ABSTRACT

We use a free–energy minimization approach to describe in simple and clear physical terms the secular and dynamical instabilities as well as the bifurcations along well–known sequences of rotating, self–gravitating fluid and stellar systems such as the Maclaurin spheroids, the Jacobi, Dedekind, and Riemann ellipsoids, and the fluid/stellar disks. Our approach stems from the Landau–Ginzburg theory of phase transitions. In this paper, we focus on the Maclaurin sequence of oblate spheroidal equilibria and on the effects of nonaxisymmetric, second–harmonic disturbances.

The free–energy approach has been pioneered in astrophysics by Bertin & Radicati (1976) who showed that the secular instability beyond the Maclaurin–Jacobi bifurcation can be interpreted as a second–order phase transition. We show that second–order phase transitions appear on the Maclaurin sequence also at the points of dynamical instability (bifurcation of the $x=+1$ self–adjoint Riemann sequence) and of bifurcation of the Dedekind sequence. The distinguishing characteristic of each second–order phase transition is the conservation/nonconservation of an integral of motion (a "conserved/nonconserved current") which, in effect, determines uniquely whether the transition appears or not. The secular instability beyond the Jacobi bifurcation appears only if circulation is not conserved. The secular instability at the Dedekind bifurcation appears only if angular momentum is not conserved. We show by an explicit calculation that, in the presence of dissipation agents that violate one or the other

[1] Virginia Institute for Theoretical Astronomy, Department of Astronomy, University of Virginia, P.O. Box 3818, Charlottesville, VA 22903

[2] NASA Goddard Space Flight Center, Code 665, Greenbelt, MD 20771

[3] Gauss Foundation Fellow

[4] Department of Physics & Astronomy, University of Kentucky, Lexington, KY 40506

[5] Department of Physics & Astronomy, Louisiana State University, Baton Rouge, LA 70803

conservation law, the global minimum of the free–energy function beyond the onset of secular instability belongs to the Jacobi and to the Dedekind sequence, respectively.

In the case of a "perfect" fluid which conserves both circulation and angular momentum, the "secular" phase transitions are no longer realized and the Jacobi/Dedekind bifurcation point becomes irrelevant. The Maclaurin spheroid remains at the global minimum of the free–energy function up to the bifurcation point of the $x=+1$ Riemann sequence. The $x=+1$ equilibria have lower free energy than the corresponding Maclaurin spheroids for the same values of angular momentum and circulation. Thus, a "dynamical" second–order phase transition is allowed to take place beyond this bifurcation point. This phase transition brings the spheroid, now sitting at a saddle point of the free–energy function, to the new global minimum on the $x=+1$ Riemann sequence.

Circulation is not conserved in stellar systems because the stress–tensor gradient terms that appear in the Jeans equations of motion include "viscosity–like" off–diagonal terms of the same order of magnitude as the conventional "pressure" gradient terms. For this reason, globally unstable axisymmetric stellar systems evolve toward the "stellar" Jacobi sequence on dynamical time scales. This explains why the Jacobi bifurcation is a point of *dynamical* instability in stellar systems but only a point of *secular* instability in viscous fluids.

The second–order phase transitions on the Maclaurin sequence are discussed in relation to the dynamical instability of stellar systems, the $\lambda$–transition of liquid $^4$He, the second–order phase transition in superconductivity, and the mechanism of spontaneous symmetry breaking.

## 1   INTRODUCTION

The equilibrium structure and stability characteristics of spheroidal and generally triaxial self–gravitating fluid masses with uniform density and uniform rotation and vorticity have been studied for many centuries by many distinguished researchers because of their relative mathematical simplicity and their possible applications to celestial objects. A modern and eloquent presentation of the entire subject is given by Chandrasekhar (1969) in his monograph "Ellipsoidal Figures of Equilibrium" (hereafter referred to as EFE). The best studied figures of uniformly rotating, self–gravitating, homogeneous, incompressible fluid masses in equilibrium are: the Maclaurin spheroids that are oblate spheroidal in shape (see Florides & Spyrou 1994 for a surprising recent reminder that prolate Maclaurin spheroids do not exist; see also Lamb 1932 §374); the Jacobi ellipsoids that have no vorticity viewed from a rotating frame in which the figure appears stationary; the Dedekind ellipsoids that have no rotation and are supported completely by internal motions of uniform vorticity; and the Riemann S–type ellipsoids that have rotation and vorticity vectors both aligned with a symmetry axis of the figure.

Linear stability analysis of spheroids and ellipsoids matured with Chandrasekhar's work and his use of the tensor virial method (EFE). His principal results concerning points of bifurcation (where a new sequence of equilibria branches off) and points of dynamical instability (where the structure of the system is changed drastically on a dynamical time scale by the action of exponentially growing disturbances) can be summarized as follows. (a) All Riemann sequences of S–type ellipsoids bifurcate from the dynamically stable part (see below) of the Maclaurin sequence. The Jacobi and Dedekind sequences bifurcate at a meridional eccentricity of $e$=0.81267 where either of two second–harmonic modes of oscillation may become neutral by an appropriate choice of a coordinate system. With respect to second–harmonic disturbances, Maclaurin spheroids are dynamically unstable for $e \geq 0.95289$ and secularly unstable for $e \geq 0.81267$; the Jacobi, Dedekind, and general Riemann S–type ellipsoids are dynamically stable. (b) Two equilibrium sequences of pear–shaped objects bifurcate from the Maclaurin sequence at $e$=0.89926 and $e$=0.96937; two additional equilibrium sequences of pear–shaped objects also bifurcate one from the Jacobi sequence at $e$=0.938577, $\eta$ =0.901762 (where $\eta$ is the equatorial eccentricity) and another from the Dedekind sequence at $e$=0.936597, $\eta$=0.897345; at each of these points a third–harmonic mode of oscillation becomes neutral. With respect to third–harmonic disturbances, the Maclaurin spheroids are dynamically unstable for $e \geq 0.96696$ and secularly unstable for $e \geq 0.89926$; the Jacobi ellipsoids stand out among the above mentioned configurations in that the point of third–harmonic dynamical instability occurs exactly at the point of bifurcation of pear–shaped objects ($e$=0.938577, $\eta$ =0.901762).

We shall return to this interesting coincidence briefly in the second and mainly in the third paper of this series (hereafter referred to as Paper II and Paper III, respectively). Specifically, we plan to continue our analysis of bifurcation points by considering axisymmetric and fourth–harmonic disturbances in Paper II. Paper III will be devoted to the third–harmonic points of the Maclaurin and the Jacobi sequences and their relevance to the fission problem (e.g. Lebovitz 1972, 1987).

In this paper, we consider only second–harmonic disturbances applied to Maclaurin spheroids because such perturbations have been thoroughly investigated over the past twenty years in both fluid and stellar systems although several fundamental questions still remain unanswered (see §4 below). Furthermore, there is no essential loss of generality in considering only the Maclaurin spheroids and it is in this idealized context that our interpretations of results become most transparent and easy to realize physically.

Equilibrium sequences of Riemann S–type ellipsoids can be depicted as curved lines on the axes ratio plane ($b/a, c/a$) where $a, b, c$ are the three principal axes along the $X, Y, Z$ axes of the coordinate frame in which the figures appear to be stationary (EFE). The sequences are bounded in that plane by the stable part of the Maclaurin sequence and by two curves that begin at the "needle" $b/a = c/a = 0$ and terminate on the Maclaurin sequence at the point of dynamical instability $b/a = 1$, $c/a = 0.30333$ and at the nonrotating sphere $b/a = c/a = 1$ (EFE; Lebovitz 1972). These two bounding sequences are known as the lower and upper self–adjoint sequences with $x$=+1 and with $x = -1$, respectively.

The Riemann ellipsoids have four integrals of motion, the mass, the angular momentum, the circulation, and the Jacobi integral (EFE; Binney & Tremaine 1987). The Jacobi integral is the sum of a "kinetic energy" term associated with expansion/contraction of the axes during evolution and a "potential energy" term which we shall call the free–energy function for reasons that will be made clear below (see also Tohline & Christodoulou 1988). As Chandrasekhar describes, Riemann in his 1860 paper not only introduced and studied triaxial ellipsoids for stability but also realized the importance of the free–energy function from which he showed that the Maclaurin spheroids become unstable at $e=0.95289$. Riemann's result implies in effect that minima of the free–energy function for variations of the axes that conserve mass, angular momentum, and circulation occur along the Maclaurin sequence only up to $e=0.95289$ and then on to the lower self–adjoint sequence with $x=+1$ where stability is thus necessarily transferred. The precise free–energy minimization that leads to this conclusion can be found in §53 of EFE. We note, however, that the same result can be obtained from physical arguments which are also discussed in §53 of EFE. We clarify this point in §3.1 below.

With respect to the other critical point ($e=0.81267$ corresponding to the onset of secular instability), an important discovery was made by Bertin & Radicati (1976) who showed that the Jacobi bifurcation point on the Maclaurin sequence can be understood by Landau–Ginzburg theory of second–order phase transitions (Landau & Lifshitz 1986) with the equatorial eccentricity $\eta$ playing the role of the order parameter (a quantity that vanishes in the state of higher symmetry, i.e., on the Maclaurin sequence). Hachisu & Eriguchi (1983) have since expanded this work and have suggested that more points can be thought as phase transition points of first and third order. We shall return to these results of Hachisu & Eriguchi in Paper II. It is also clear that the free–energy approach was also understood and appreciated by Lyttleton (1953) who gave a general discussion of the stability of Maclaurin spheroids and Jacobi ellipsoids based on properties of the energy function of these systems but, then, continued his analysis using only the equations of motion.

Our work was motivated by the above intriguing results of Riemann (EFE), Bertin & Radicati (1976), and Hachisu & Eriguchi (1983). It is well–understood that the minima of the free–energy function determine *stable* equilibria available to an evolving system. Similarly, the maxima of the free–energy function determine *unstable* equilibria. In §3, we show that minimization of the free–energy function (cf. §53 in EFE) under the additional assumption that circulation is not conserved (or, alternatively, angular momentum is not conserved) leads to the Jacobi sequence (or, alternatively, to the Dedekind sequence) of ellipsoidal equilibria. (Obtaining the Jacobi sequence and, in turn, the Dedekind sequence as minimum energy states is yet another manifestation of Dedekind's theorem; see EFE.) Since minimization of the free energy isolates the equilibrium sequences along which stability is transferred for both secular and dynamical instability, we are led to the conclusion that points of bifurcation and of dynamical instability along any equilibrium sequence can be understood from the energetics of the perturbed systems that subsequently evolve out of equilibrium. Phase–transition theory conveniently allows one to follow the energetics of non–equilibrium systems and there also lies the importance of the result obtained by Bertin & Radicati (1976).

We should emphasize that the phase transition of a physical system is an intrinsically *non–equilibrium* process and it can only be studied in detail by considering the behavior of a system out of equilibrium. For example, the equilibrium study of Bertin & Radicati (1976) successfully identified the Jacobi bifurcation point with the appearance of a second–order phase transition but could not determine whether the transition is *allowed* or *forbidden*. For such an identification, one needs to study nonequilibrium states and there the question of (non)–conservation of circulation is immediately manifested. Along very similar lines, Hachisu & Eriguchi (1983) have identified more phase transition points by considering only equilibrium sequences but they did not realize that the third–order phase transition at the bifurcation point of the one–ring sequence is *forbidden* on energetic grounds. This conclusion follows from their results by considering nonequilibrium states (see Paper II for details). The nonequilibrium nature of phase transitions is true for first–order (Frenkel 1955; Huang 1963; Abraham 1974) as well as for second–order phase transitions (London 1950; Landau & Lifshitz 1986). We have stressed this point previously ourselves in the context of a first–order phase transition during protostellar collapse and star formation (Tohline 1985; Tohline, Bodenheimer, & Christodoulou 1987; Tohline & Christodoulou 1988; Christodoulou & Tohline 1990; Christodoulou, Sasselov, & Tohline 1993). This is the same first–order phase transition identified by Hachisu & Eriguchi (1983) in the vicinity of the one–ring sequence (see Paper II for more details).

In this work, we employ the fully nonlinear techniques of our previous work concerning phase transitions in order to study in and out of equilibrium incompressible Riemann S-type ellipsoids (EFE) and the corresponding compressible Riemann disks (Weinberg & Tremaine 1983; Weinberg 1983). We show that calculation of the free energy under non–equilibrium conditions in evolving disks and ellipsoids leads to the following physical interpretation of the second–harmonic points of bifurcation and of dynamical instability on the Maclaurin sequence: Under the assumption of nonconservation of circulation, the Jacobi bifurcation point is singled out because a second–order phase transition appears beyond that point. This phase transition toward the Jacobi sequence appears only in the presence of a viscous, dissipative mechanism which destroys vorticity and thereby results in nonconservation of circulation. Circulation thus varies on a viscous time scale which justifies the name "secular instability" given to this particular phase transition. This second–order phase transition is not realized in an inviscid, "perfect" fluid. Under the assumption of nonconservation of angular momentum in an inviscid fluid (as in the case of gravitational radiation or any other agent that causes angular momentum losses but conserves circulation), another second–order phase transition appears at the same point and leads to the Dedekind sequence. On the other hand, an "ideal" Maclaurin spheroid that conserves both its angular momentum and its circulation exactly continues to be on the global minimum of the free energy past the Jacobi bifurcation point and up to the bifurcation point of the $x=+1$ self–adjoint Riemann sequence. Beyond that point, the Maclaurin spheroid finds itself on a saddle point of the free energy while a new global minimum, that belongs to the $x=+1$ sequence, appears. Therefore, a third second–order phase transition appears and is naturally allowed to take place at once toward the $x=+1$ sequence because both angular momentum and circulation are automatically conserved between the Maclaurin and the $x=+1$ sequence. Thus, this second–order phase transition proceeds on a dynamical time scale which justifies the name "dynamical instability" given to this particular phase transition.

The remainder of the paper is organized as follows. In §2, we summarize the equilib-rium properties of Riemann ellipsoidal systems and we present the free–energy function for non–equilibrium evolution driven by second–harmonic disturbances. In §3, we show that it is (non)–conservation of circulation that determines both the appearance and the characteristic time scales of the two well–known instabilities on the Maclaurin sequence. We also examine the consequences of nonconservation of angular momentum when circulation is strictly conserved in Maclaurin spheroids and when the adiabatic invariants are conserved in stellar elliptical disks (Hunter 1974). In §4, we address the secular and dynamical instabilities of Riemann ellipsoidal systems in light of our results. We also discuss the implications of nonconservation of circulation for the dynamical evolution of *stellar* disks and for their global stability to second–harmonic perturbations. Finally, we discuss the second–order phase transitions on the Maclaurin sequence in relation to the $\lambda$–transition of liquid $^4$He, the second–order phase transition in superconductivity, and the mechanism of spontaneous symmetry breaking.

In an Appendix to this paper, we formulate a helpful analogy between the phase transitions discussed in this paper and the thermodynamical phase transitions.

## 2   Riemann Ellipsoidal Equilibria

### 2.1   Riemann S-type Ellipsoids

As a class, the generally triaxial Riemann S-type ellipsoids include all the above mentioned equilibrium sequences (Maclaurin, Jacobi, Dedekind, upper and lower self–adjoint). All S-type ellipsoids have their angular momentum and vorticity vectors aligned with one of the principal axes of the figure. Riemann sequences of S-type equilibria are characterized by specific relations between the three principal axes $a, b, c$ (along the axes $X, Y, Z$, respectively, of the coordinate system in which each figure appears stationary) and the rotation frequency $\Omega$ of the coordinate system ; they are defined in EFE by the dimensionless parameter

$$f \equiv \frac{\zeta}{\Omega} = \text{constant}, \tag{2.1}$$

where $\zeta$ is the vorticity in the rotating coordinate system. All S-type Riemann sequences bifurcate from the dynamically stable part of the Maclaurin sequence of oblate spheroids. The Jacobi and Dedekind sequences have $f = 0$ and $f = \pm\infty$, respectively.

The relevant equilibrium equations are somewhat simpler when they are expressed in terms of another dimensionless parameter $x$ (in place of $f$) such that

$$x = \frac{ab}{a^2 + b^2} f, \tag{2.2}$$

and for $a \geq b, c$ they read (EFE)

$$x^2 + \frac{2abB_{12}}{c^2 A_3 - a^2 b^2 A_{12}} x + 1 = 0, \tag{2.3}$$

$$\frac{\Omega^2}{\pi G \rho} = \frac{2B_{12}}{1 + x^2}, \tag{2.4}$$

where $\rho$ is the density of the homogeneous incompressible fluid and $G$ is the gravitational constant. The index symbols $A_i$, $A_{ij}$, and $B_{ij}$ $(i, j = 1, 2, 3)$ are functions of the principal axes $a, b, c$ of the ellipsoid and are given in EFE. Additional equilibrium constraints on the principal axes, that secure the reality of $x$ derived from equation (2.3), are also given in EFE.

The angular momentum $L$, circulation $C$, and free energy $E$ are also given in EFE. For our purposes, we choose units such that $\pi \rho = G = M/5 = 1$, where $M \equiv 4\pi \rho abc/3$ is the mass of the ellipsoid, and we adopt the following expressions that are equivalent to those given in EFE:

$$L = \Omega(a^2 + b^2 + 2abx), \tag{2.5}$$

$$C = -\Omega\Big[2ab + (a^2 + b^2)x\Big], \tag{2.6}$$

and

$$E = \frac{1}{2}\Omega^2\Big[(a + bx)^2 + (b + ax)^2\Big] - 2I. \tag{2.7}$$

These equations are obtained by direct integrations in the inertial system of coordinates. Angular momentum and circulation are defined to be positive and negative, respectively, along the $+Z$ axis. The first term in equation (2.7) represents the total kinetic energy per unit mass $T$ while the second term is minus twice the specific moment of inertia $I$ and represents the total gravitational potential energy per unit mass $W$ for the adopted units. The specific moment of inertia of a Riemann ellipsoid is defined by

$$I \equiv A_1 a^2 + A_2 b^2 + A_3 c^2. \tag{2.8}$$

Because of the particular choice of units, our calculations must be supplemented by an equation of mass conservation of the form

$$abc = \frac{15}{4}. \tag{2.9}$$

This equation follows from the definition of $M$ for $\pi \rho = 1$ and $M = 5$.

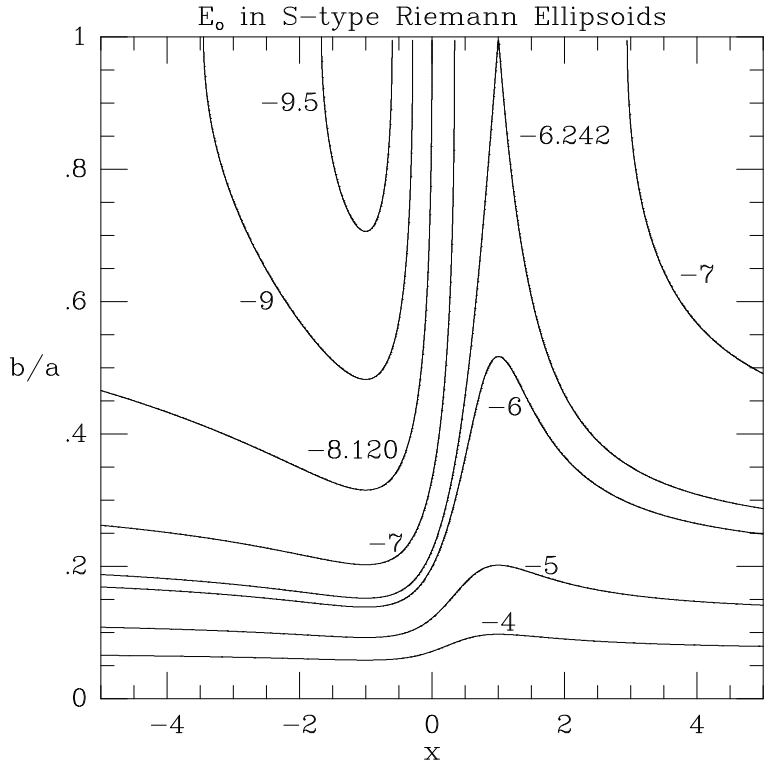

FIGURE 1. *Contours of the equilibrium free energy $E_o$ of Riemann S–type ellipsoids are plotted in the $(x, b/a)$ plane. According to their $x$–values, the Maclaurin spheroids along the axis $b/a = 1$ appear as first members of Riemann sequences of ellipsoids with $f = 2x$ [cf. equations (2.1) and (2.2)]. The Jacobi sequence has $x=0$ and the lower self–adjoint Riemann sequence has $x=+1$.*

The free energy $E = E_o$ in equilibrium can be determined from equation (2.7) as a function of the axis ratio $b/a$ and the parameter $x$ after using equations (2.3), (2.4), and (2.9) to determine $a, b, c,$ and $\Omega$. A contour plot of $E_o(x, b/a)$ is shown in Figure 1.

## 2.2   Elliptical Riemann Disks

The compressible elliptical Riemann disks have many common characteristics with the corresponding S-type ellipsoids (Weinberg & Tremaine 1983). For example, they all bifurcate from the dynamically stable part of the Maclaurin sequence of circular disks and all members are dynamically stable to second–harmonic disturbances (Weinberg 1983). The free–energy function in equilibrium is yet another common property (it is very similar to Figure 1) reflecting the dynamical correspondence between Riemann disks and ellipsoids. In fact, the equations of Weinberg & Tremaine (1983) for angular momentum, circulation, and energy can be cast into the forms given above for Riemann ellipsoids (with the exceptions of the potential energy term $W$ which no longer depends on the axis ratio $c/a$ and the internal energy term $U$ of the compressible system). We conclude that the dynamical evolution of Riemann disks out of equilibrium parallels closely that of S-type ellipsoids.

Weinberg & Tremaine (1983) have plotted two–dimensional contour maps of the angular momentum $L$, circulation $C$, and energy $E$ for Riemann disks and discussed the secular evolution of the disks (in the direction of decreasing $E$) as trajectories on these maps under the action of viscosity (only $L$ is conserved), gravitational radiation (only $C$ is conserved), and gravitational torques in a galaxy environment. Their findings are consistent with our results described in §§3.2, 3.3 below. The results in §3 are therefore valid for Riemann disks as well as for ellipsoids although no explicit reference is made to disks in that section prior to §3.3.

## 3   FREE ENERGY OUT OF EQUILIBRIUM

We calculate the free–energy function $E$ for evolving, generally out-of-equilibrium, ellipsoids using only equations (2.5)–(2.9), i.e., after relaxing the assumption of equilibrium that was expressed in §2 through equations (2.3) and (2.4). In general, $E$ is a function of five variables $a, b, c, x,$ and $\Omega$. One variable is eliminated from equation (2.9). One or two more variables are further eliminated from equations (2.5), (2.6) depending on the assumption that either $L$ or $C$ or both would be conserved during evolution. We analyze each case separately below.

### 3.1   Both $L$ and $C$ are Conserved

This case is treated in §53 of EFE. Eliminating $\Omega$ and $x$ between equations (2.5)–(2.7) one obtains the free–energy function

$$E = \frac{K_1^2}{(a-b)^2} + \frac{K_2^2}{(a+b)^2} - 2I, \tag{3.1}$$

where the two new integrals of motion are defined by

$$K_1 \equiv \frac{1}{2}(L+C) = \frac{1}{2}\Omega(a-b)^2(1-x), \tag{3.2}$$

$$K_2 \equiv \frac{1}{2}(L-C) = \frac{1}{2}\Omega(a+b)^2(1+x). \tag{3.3}$$

The first term in equation (3.1) becomes singular if at any time during evolution the object becomes momentarily axisymmetric with $a = b$. One therefore sets $K_1 \equiv 0$ in equation (3.1) and proceeds to minimize $E$ as a function of two variables $a, b$ for constant $K_2 \equiv \Omega(a+b)^2$, i.e., for $x=+1$. Axis $c$ is a dependent variable through equation (2.9). Energy minimization produces extrema only along the Maclaurin sequence and along the $x=+1$ self–adjoint sequence. In contrast to this calculation, we see from equation (3.2) that the condition $K_1 \equiv 0$, which must hold at all times, is satisfied on the Maclaurin sequence ($a = b$) and away from the Maclaurin sequence only if $x=+1$. The same result is thus recovered by simply considering equation (3.2) along with the physical requirement that $K_1 \equiv 0$ (i.e., $L = -C$ and $x=+1$) at all times.

The simplest way to determine whether the Maclaurin spheroids and the $x=+1$ ellipsoids are maxima or minima of the free–energy function is to plot $E$ from equation (3.1) as a function of $b/a, c/a$ along with equations (2.8), (2.9) and for $K_1 \equiv 0$, $K_2$=constant. Each value of $K_2$ corresponds to a Maclaurin spheroid which turns out to be the only minimum energy state if its meridional eccentricity is $e < 0.9528867$ (corresponding to $L \equiv K_2 < 7.094887$). Beyond $e = 0.9528867$, a new global minimum appears while the Maclaurin spheroid finds itself on a saddle point. This is illustrated in Figure 2 where $E(c/a, b/a)$ is plotted for $e = 0.96$ corresponding to a Maclaurin spheroid with $c/a$=0.28, $L = K_2 = 7.41288$, and $E_o = -6.03195$. This point is a saddle point in Figure 2. To within the numerical accuracy of this calculation, the global minimum $E = -6.03745$ is found at $b/a$=0.546, $c/a$=0.219 in Figure 2. It can be easily verified from equations (2.3), (2.4), and (2.9) that the new energy minimum does indeed belong to the $x=+1$ Riemann sequence.

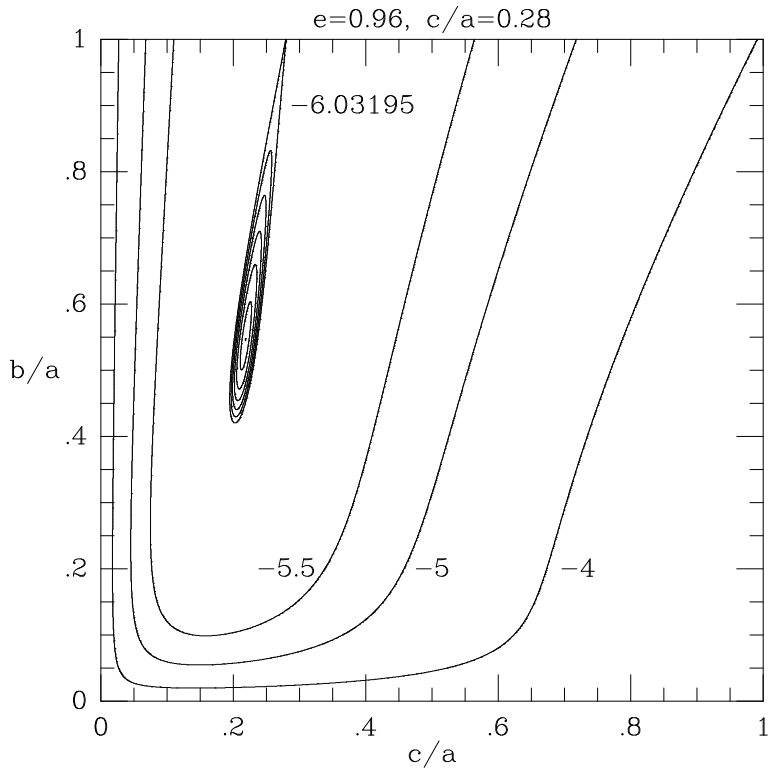

FIGURE 2. *Contours of the free–energy function $E(x = 1)$ of out-of-equilibrium triaxial ellipsoids are plotted in the $(c/a, b/a)$ plane using equation (3.1) with $K_1 \equiv 0$ (or, equivalently, $x = +1$) and $L \equiv K_2$=7.41288. This value of $L$ corresponds to a Maclaurin spheroid slightly above the point of dynamical instability with $e$=0.96, $c/a$=0.28, and $E_o = -6.03195$. Unmarked contours are drawn at intervals of $\Delta E = 10^{-3}$ between $E = -6.03695$ and $E = -6.03295$. The Maclaurin spheroid sits on a saddle point while a global minimum with $E_o = -6.03745$ (indicated by a dot inside the unmarked contours) exists at $b/a$=0.546, $c/a$=0.219. Under the assumption of exact conservation of circulation a second–order phase transition takes place on a dynamical time scale from the higher to the lower energy state following the slope highlighted by the unmarked contours.*

Figure 2 is also an evolutionary diagram. Since both $L$ and $C \equiv -L$ are conserved in this diagram, the Maclaurin spheroid will evolve on a dynamical time scale from its original position to the new energy minimum following the direction of steepest descent. This direction is clearly seen in Figure 2 as a ridge connecting the two extrema of the free energy.

### 3.2   Only $L$ is Conserved

Eliminating $\Omega$ between equations (2.5) and (2.7) we obtain the free–energy function

$$E = \frac{L^2}{2} \frac{(a + bx)^2 + (b + ax)^2}{(a^2 + b^2 + 2abx)^2} - 2I. \tag{3.4}$$

Out of equilibrium, $E$ is now a function of three variables since one of the axes can be eliminated from equation (2.9). However, since $x$ appears only in the kinetic energy term of equation (3.4), we can minimize $E$ first with respect to $x$. We find that

$$\frac{\partial E}{\partial x} = \frac{L^2(a^2 - b^2)^2 x}{(a^2 + b^2 + 2abx)^3}. \tag{3.5}$$

Letting $\partial E/\partial x = 0$ we find for $a \neq b$ that $x = 0$. (The solutions $x \to \pm\infty$ are rejected because all higher derivatives are identically zero.) Since $\partial^2 E/\partial x^2 > 0$, $x = 0$ may only represent a minimum of the free energy. We can conclude at this point from equation (3.5) that we have obtained the Maclaurin sequence ($a = b$) and the Jacobi sequence ($x=0$) as the only possible extrema of the free energy. A formal minimization of the function

$$E(x = 0) = \frac{L^2}{2(a^2 + b^2)} - 2I, \tag{3.6}$$

along with equations (2.8), (2.9), performed in the spirit of §53 in EFE, confirms this conclusion. It leads to the triangle of equalities

$$\frac{L^2}{2(a^2 + b^2)^2} = A_1 - \frac{c^2}{a^2}A_3 = A_2 - \frac{c^2}{b^2}A_3, \tag{3.7}$$

which describe the Maclaurin spheroids for $a = b$ and the Jacobi ellipsoids for $a \neq b$ (cf. §32 and §39 in EFE, respectively).

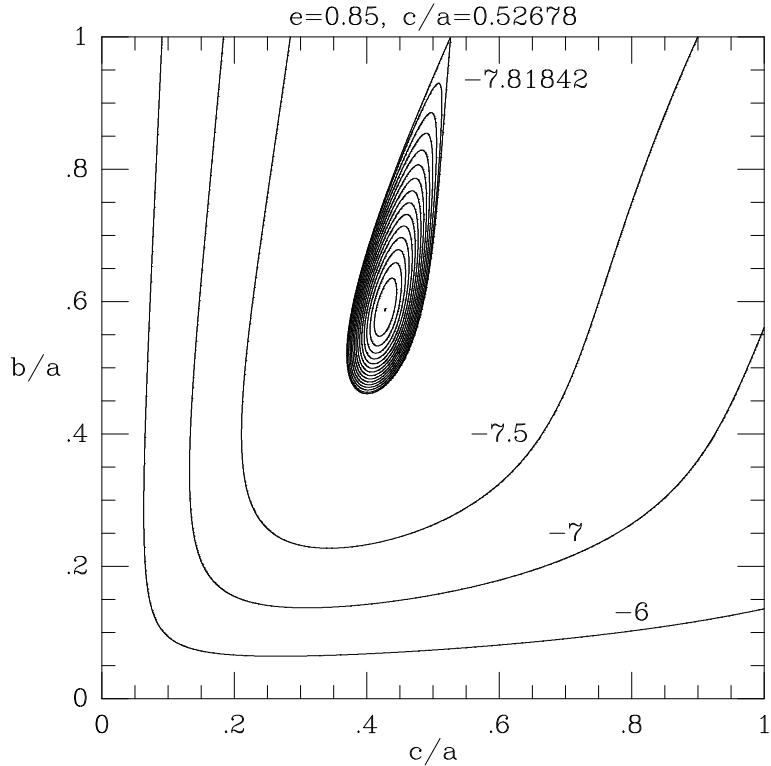

FIGURE 3. Contours of the free–energy function $E(x = 0)$ of out-of-equilibrium triaxial ellipsoids are plotted in the $(c/a, b/a)$ plane using equation (3.6) with $L=4.71488$. This value of $L$ corresponds to a Maclaurin spheroid above the point of secular instability with $e=0.85$, $c/a=0.52678$, and $E_o = -7.81842$. Unmarked contours are drawn at intervals of $\Delta E = 10^{-3}$ between $E = -7.832$ and $E = -7.819$. The Maclaurin spheroid sits on a saddle point while a global minimum with $E_o = -7.83300$ (indicated by a dot inside the unmarked contours) exists at $b/a=0.588$, $c/a=0.428$. Under the action of viscous dissipation circulation is not conserved and a second–order phase transition takes place on a viscous time scale from the higher to the lower energy state following the slope highlighted by the unmarked contours.

The simplest way to determine whether the Maclaurin spheroids and the Jacobi ellipsoids are maxima or minima of the free–energy function is again to plot $E(x = 0)$ from equation (3.6) as a function of $b/a, c/a$ along with equations (2.8), (2.9) and for $L$=constant. Each value of $L$ corresponds to a Maclaurin spheroid which turns out to be the only minimum energy state if its meridional eccentricity is $e < 0.8126700$ (corresponding to $L < 4.232964$). Beyond $e = 0.8126700$, a new global minimum appears while the Maclaurin spheroid finds itself on a saddle point. This is illustrated in Figure 3 where $E(x = 0)$ is plotted for $e = 0.85$ corresponding to a Maclaurin spheroid with $c/a=0.52678$, $L = 4.71488$, and $E_o = -7.81842$. This point is a saddle point in Figure 3. To within the numerical accuracy of this calculation, the global minimum $E = -7.83300$ is found at $b/a=0.588$, $c/a=0.428$ in Figure 3 and does belong to the Jacobi sequence since it satisfies the triangle of equalities (3.7).

Like Figure 2 above, Figure 3 is also an evolutionary diagram. Since only $L$ is conserved in Figure 3, however, the Maclaurin spheroid will evolve on a viscous time scale characteristic of the variation of circulation induced by viscous dissipation. The direction of steepest descent along which evolution proceeds is clearly seen in Figure 3 as a ridge connecting the two extrema of the free energy.

### 3.3   Only C is Conserved

Eliminating $\Omega$ between equations (2.6) and (2.7) we obtain the free–energy function

$$E = \frac{C^2}{2} \frac{(a+bx)^2 + (b+ax)^2}{[2ab + (a^2+b^2)x]^2} - 2I. \tag{3.8}$$

Extrema of the free–energy function with respect to $x$ may now be obtained by first transforming $x \to x' \equiv 1/x$ in equation (3.8). Differentiating $E$ with respect to $x'$ we obtain equation (3.5) with $x'$ in place of $x$. The condition $\partial E/\partial x' = 0$ leads to the solution $x' = 0$ for which $x \to \pm\infty$ and

$$E(x \to \pm\infty) = \frac{C^2}{2(a^2+b^2)} - 2I. \tag{3.9}$$

This equation has the same form as equation (3.6) but $C$ has taken the place of $L$. Minimizing $E(x \to \pm\infty)$ with respect to variations of axes that obey equation (2.9), we obtain the triangle of equalities (3.7) with $C$ in place of $L$. These equations define the Dedekind sequence for $a \neq b$. This result was expected since $x \to \pm\infty$ only on that sequence. Furthermore, this result was also expected as a consequence of Dedekind's theorem (see §28 in EFE) because of the transposition between $L$ and $C$ and between $x$ and $x' = 1/x$ in equations (3.4) and (3.8).

This case, where only circulation is conserved, may be applicable to systems that evolve by emitting gravitational radiation and also to galactic stellar bars only if they lose angular momentum due to external gravitational torques (Weinberg & Tremaine 1983). Such systems will undergo a second–order phase transition toward a minimum energy state that belongs to the Dedekind sequence on a time scale characteristic of the rate of angular momentum loss.

The work of Hunter (1974) reveals another related example. Hunter considered the evolution of stellar, generally elliptical, "Freeman" disks as they suffer slow mass loss. During evolution, two integrals of motion, called the adiabatic invariants, are conserved. Mass loss, however, causes angular momentum loss with the following consequences. Elliptical disks generally tend to become circular with time although a certain subgroup tend to become nonrotating with $b/a \lesssim 0.4$ in the final $\Omega = 0$ state (see Figure 4 in Hunter 1974). More importantly for our purposes, a subgroup of circular disks with low rotation frequencies $\Omega$ of the coordinate system tend to become elliptical and nonrotating with $b/a \gtrsim 0.4$ in the final $\Omega = 0$ state. These circular stellar disks undergo a second–order phase transition toward the corresponding "stellar–disk Dedekind sequence." This phase transition occurs below $T/|W| = 0.12860$ and should not be confused with the dynamical instability above the exact same value of $T/|W|$ discovered by Kalnajs (1972) at high rotation frequencies $\Omega$. The dynamical instability does not depend on conservation of the adiabatic invariants (only $L$ is conserved) and represents a second–order phase transition toward the "stellar–disk Jacobi sequence." We have been able to show this by applying the energy minimization technique of §3.2 to the energy function given by Freeman (1966) for the stellar disks.

## 4  DISCUSSION

We have used a free–energy minimization principle that derives from the thermodynamical theory of phase transitions in order to describe a simple physical picture that explains nonaxisymmetric secular and dynamical instabilities of second–harmonic perturbations in rotating self–gravitating fluids. The incompressible Maclaurin spheroids and, in general, triaxial Riemann ellipsoids have served as an ideal "laboratory" in our study. Our results are, however, directly applicable to compressible Riemann disks as well as to two–dimensional stellar disks.

Both secular and dynamical second–harmonic instabilities can be interpreted as second–order phase transitions with the equatorial eccentricity of the nonaxisymmetric state (low–symmetry phase) playing the role of the order parameter. This result was previously known only in the case of the secular second–harmonic instability that appears on the Maclaurin sequence beyond the Jacobi bifurcation point in the presence of viscous dissipation (Bertin & Radicati 1976). However, the connection between such a phase transition and the conservation laws had not been appreciated in the past. We have shown that whether a phase transition takes place toward the Jacobi or the Dedekind sequence is decided by the conservation laws. We have also shown that the dynamical second–harmonic instability that appears on the Maclaurin sequence beyond the bifurcation point of the $x=+1$ sequence represents yet another second–order phase transition with the same order parameter and the same type of symmetry breaking, i.e. the axisymmetric Maclaurin spheroid becomes a triaxial Riemann ellipsoid.

The existence of all three second–order phase transitions depends crucially on whether the angular momentum and the circulation integrals are exactly conserved or not and on how quickly they may be varying during evolution (see also §3.3 for the analogous case in stellar disks). The phase transition at the Jacobi(Dedekind) bifurcation point of meridional eccentricity $e=0.81267$ is allowed to take place only if angular momentum (circulation) is conserved and circulation (angular momentum) decreases due to a viscous mechanism (angular momentum losses). These phase transitions take place on the same time scales that vorticity is destroyed by the dissipative mechanism and angular momentum is lost from the system, respectively. The end–point of each transition is a lower energy state that belongs to the Jacobi or to the Dedekind sequence of equilibrium ellipsoids, respectively (see Figure 3 above). The phase transitions at $e=0.81267$ are also applicable to stellar disk systems. We discuss this case in detail in §4.2 below (but see also §3.3).

In the ideal case of absence of viscosity and gravitational radiation, a fluid is "perfect" and nonrelativistic and conserves both its angular momentum and its circulation exactly at all times. As soon as the free–energy function is further constrained by both conservation laws, the above phase transitions disappear and the Jacobi/Dedekind bifurcation point becomes irrelevant. Consequently, the Maclaurin spheroid continues to be on the global minimum of the free–energy function past this bifurcation point and up to $e{=}0.95289$ where another sequence of Riemann equilibria, the $x{=}{+}1$ self–adjoint sequence, bifurcates. The ellipsoids of this sequence have lower total energy and exactly the same angular momentum $L$ and circulation $C \equiv -L$ compared to the spheroids of the Maclaurin sequence past the bifurcation point (see Figure 4 below). Under the circumstances, another second–order phase transition then appears past the point $e{=}0.95289$ and is automatically allowed to take place because all conservation laws are satisfied during the transition. Since this second–order phase transition does not require assistance from dissipation of any kind, it proceeds on a dynamical time scale. The Maclaurin spheroid which finds itself on a saddle point of the free energy past the point $e{=}0.95289$ is free to "roll down" toward a new energy minimum that belongs to the self–adjoint Riemann sequence with $x{=}{+}1$ (see Figure 2 above).

If neither circulation nor angular momentum is conserved during evolution, then all systems (fluid and stellar) end up spherical and nonrotating because this is the only available state with no vorticity and no angular momentum. For the Riemann S–type ellipsoids of Figure 1 above, this state lies at the intersection of the Maclaurin sequence ($b/a = 1$) and the upper self–adjoint Riemann sequence with $x = -1$. One can see from Figure 1 that this point represents the lowest possible energy state among all Riemann S–type ellipsoids.

### 4.1 Bifurcations and Second–Order Phase Transitions

In contrast to *smoothly* bifurcating sequences of axisymmetric equilibria such as the one–ring sequence, the Jacobi and the $x{=}{+}1$ sequences bifurcate *abruptly* from the Maclaurin sequence in the angular momentum–rotation frequency $(L, \Omega)$ plane (see Hachisu & Eriguchi 1983 and references therein). Therefore, it is not surprising that abrupt bifurcations of nonaxisymmetric sequences and axisymmetry breaking are associated with second–order phase transitions while smooth bifurcations in the $(L, \Omega)$ plane of equilibrium sequences that preserve axisymmetry are associated with third–order phase transitions (Bertin & Radicati 1976; Hachisu & Eriguchi 1983). We shall discuss this issue in more detail in Paper II where we shall see that such third–order phase transitions are not energetically allowed.

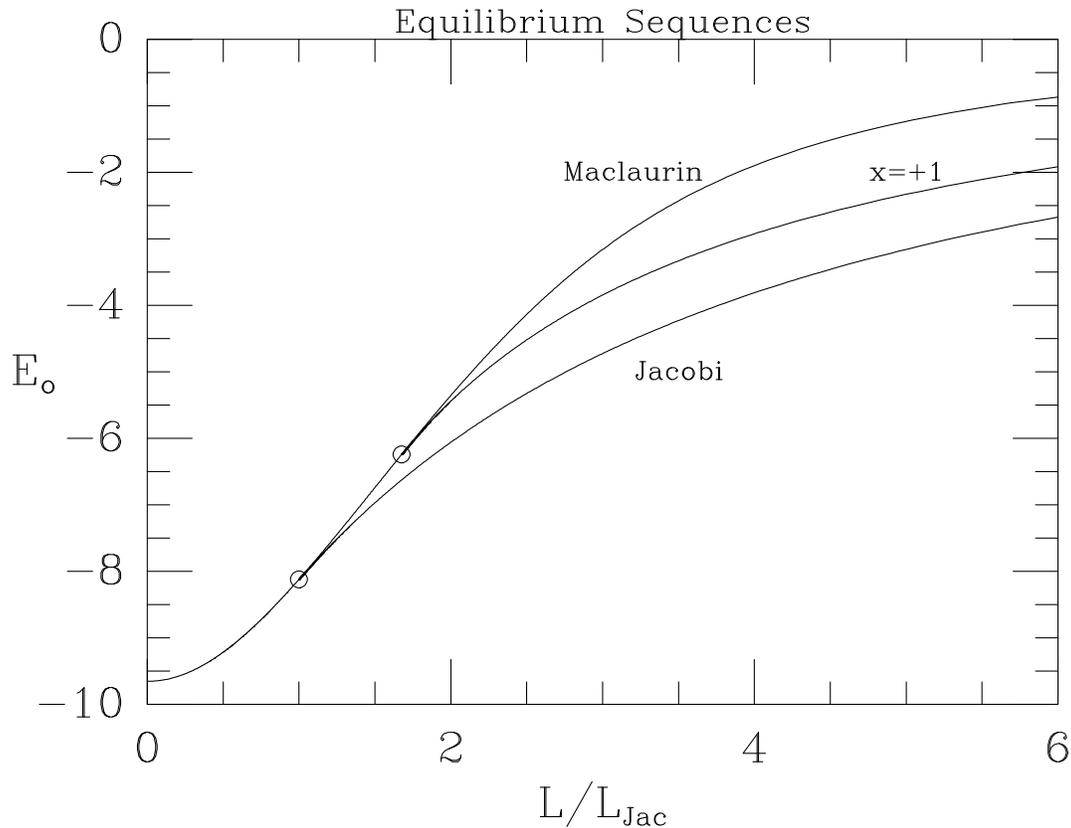

FIGURE 4. *The Maclaurin, Jacobi, and $x=+1$ equilibrium sequences are plotted in the angular momentum ($L$)–energy ($E_o$) plane. Angular momentum values are normalized to the corresponding value $L_{Jac} = 4.232964$ at the Jacobi bifurcation point. This point with $E_o = -8.119645$ and the bifurcation point of the $x=+1$ self–adjoint Riemann sequence at $L/L_{Jac} = 1.676104$ with $E_o = -6.241989$ are indicated by open circles. Unlike the corresponding diagram in the $(L, \Omega)$ plane, this plot shows the Jacobi and the $x=+1$ sequences bifurcating smoothly from the Maclaurin sequence. Smooth bifurcations indicate continuity of the first derivative ("specific entropy") of the "free energy" $E$ with respect to "temperature" $1/L$ and, therefore, possibly allowed second–order phase transitions. [Accordingly, abrupt bifurcations of the same sequences in the $(L, \Omega)$ plane indicate discontinuity of the corresponding second derivative ("specific heat") and thus again possibly allowed second–order phase transitions.]*

Here, we only note that the Jacobi and the $x=+1$ sequences bifurcate smoothly from the Maclaurin sequence in the angular momentum–energy $(L, E_o)$ plane. This is shown in Figure 4 where the equilibrium energy $E_o$ is plotted versus $L/L_{Jac}$, where $L_{Jac}=4.232964$ is the angular momentum at the Jacobi bifurcation point. The $x=+1$ sequence bifurcates smoothly at $L/L_{Jac}=1.676104$. This behavior is explained simply in analogy to thermodynamical second–order phase transitions. In a thermodynamical second–order phase transition, the specific entropy (minus the first derivative of $E$ with respect to temperature) is continuous across the transition point while the specific heat (second derivative of $E$ with respect to temperature) is discontinuous. In our case, the *inverse* of the angular momentum $1/L$ plays the role of temperature $\theta$. (Using $\theta$ for the temperature avoids confusion with the kinetic energy $T$.) In an "ideal dynamical system" analogous to an ideal gas the specific moment of inertia $I$ plays the role of specific heat $c_v$ and the *inverse* of the rotation frequency $1/\Omega$ plays the role of the internal energy $U$. This identification follows from a dimensional comparison between the dynamical equation $L = I\Omega$ and the thermodynamical equation $U = c_v\theta$ of an ideal gas.

The phase transitions at $e=0.81267$ and at $e=0.95289$ are "nonideal" in the sense that $I$ varies during each transition. However, based on the above analogy, we can still understand why these transitions are both of second order. The slopes $dE_o/dL$ of the sequences at the corresponding bifurcation points in the $(L, E_o)$ plane of Figure 4 are continuous (smooth bifurcations) while the corresponding slopes $d\Omega/dL$ in the $(L, \Omega)$ plane are discontinuous (abrupt bifurcations). This means, respectively, that the "specific entropy" is a continuous function but the "specific heat" has a finite discontinuity at the transition point, just as in a thermodynamical second–order phase transition.

The discussions in this subsection as well as in Paper II may be better understood with the help of a formal analogy between our dynamical phase transitions and the thermodynamical phase transitions. We summarize such an analogy in the Appendix.

### 4.2 The Global Stability of Stellar Systems

Lacking a solid physical foundation, the Ostriker–Peebles (1973; hereafter referred to as OP) stability criterion has been controversial since its discovery. This criterion, discovered in numerical $N$–body simulations, predicts the onset of dynamical second–harmonic instability in rotating, self–gravitating particle disks with ratios of rotational kinetic energy $T$ to gravitational potential energy $W$ in excess of $T/|W| \approx 0.14$.

Several counter–examples have been proposed where the criterion fails to predict the onset of instability or does predict instability although none exists (Toomre 1988). The simplest counter–example against the universality of the criterion has been given by Hunter (1974) and Tremaine (1976) in the case of two–dimensional stellar elliptical disks. As the axis ratio $b/a \to 0$ in maximally rotating disks ("needles"), the ratio $T/|W| \to 1/2$ (see Figure 3 in Hunter 1974 for an explicit demonstration) and so the OP criterion predicts instability. Linear stability analysis, however, shows that these "needles" are stable against second–harmonic perturbations (Tremaine 1976). If the OP criterion were naively applied to the secular evolution of fluid Jacobi "needles" with $b/a, c/a \to 0$ it would fail for the same reason although it would be successful in predicting the onset of secular instability at $T/|W| = 0.13753$ ($e=0.81267$) on the Maclaurin sequence. We have resolved several discrepancies that the OP criterion suffers from, including the ones discussed here. An improved stability criterion to second–harmonic perturbations will be discussed elsewhere. Here we only show that unstable stellar systems with $T/|W| \gtrsim 0.14$ evolve on a dynamical time scale while the corresponding fluid systems evolve secularly.

For a viscous fluid system, Navier's equation of motion reads (Truesdell 1954)

$$\mathbf{a} = -\nabla(Q + \Phi) - \frac{\mu}{\rho}\nabla \times (\nabla \times \mathbf{v}), \qquad (4.1)$$

while for a stellar system, the corresponding Jeans equation of motion reads (e.g. Binney & Tremaine 1987)

$$\mathbf{a} = -\nabla\Phi - \nabla \cdot \mathbf{T}, \tag{4.2}$$

where $Q$ denotes enthalpy, $\Phi$ is the potential due to the conservative gravitational force, $\mu$ is the coefficient of fluid dynamic viscosity, $\mathbf{T} = T_{ij}$ is the stress tensor in stellar systems, and the remaining symbols have their usual meaning. Without loss of generality, we have limited the discussion in this section to incompressible fluids and assumed that $\mu$ and $\rho$ are constant in the above equations.

The rate of change of circulation is given by (e.g. Tassoul 1978)

$$\frac{dC}{dt} = \frac{d}{dt}\oint_c \mathbf{v}\cdot d\mathbf{c} = \oint_c \mathbf{a}\cdot d\mathbf{c} = \int\int_S (\nabla\times\mathbf{a})\cdot d\mathbf{S}, \tag{4.3}$$

where $c$ denotes a closed material circuit and $S$ denotes the corresponding cross–sectional surface area of a vortex tube. The surface integral in the last step is obtained as usual by applying Stokes's theorem. We have thus arrived at the d'Alembert–Euler condition (Truesdell 1954)

$$\nabla\times\mathbf{a} = 0, \tag{4.4}$$

which is a necessary and sufficient condition that the circulation of every material circuit remain constant in time. Motions that satisfy this condition are called *circulation–preserving motions*.

Taking the curl of equation (4.1) we find for a viscous fluid system of vorticity $\zeta \equiv \nabla\times\mathbf{v}$ that

$$\nabla\times\mathbf{a} = -\frac{\mu}{\rho}\nabla\times(\nabla\times\zeta). \tag{4.5}$$

We see now that there is a special class of viscous fluids with $\nabla\times(\nabla\times\zeta) = 0$ in which the motions are circulation–preserving. These fluids are said to admit a *flexion–potential* (see Truesdell 1954 for details). In general, real viscous fluids do not admit a flexion–potential and circulation varies on a time scale dictated by viscosity [cf. equations (4.3) and (4.5)]. The instability induced in this way is secular.

Taking the curl of equation (4.2) we find for a stellar system that

$$\nabla\times\mathbf{a} = -\nabla\times(\nabla\cdot\mathbf{T}). \tag{4.6}$$

We see now that because of the anisotropic stress–tensor terms in the Jeans equation (4.2) and in equation (4.6) circulation is never conserved during the evolution of a stellar system. To estimate the characteristic time scale for variation of the circulation in stellar systems we write explicitly the stress–tensor gradients (e.g. Binney and Tremaine 1987)

$$\frac{\partial T_{ij}}{\partial x_j} = \frac{\partial}{\partial x_j}(\overline{v_i v_j} - \overline{v}_i\overline{v}_j), \tag{4.7}$$

where $v_i$, $v_j$ $(i,j=1,2,3)$ denote velocity components and the overline means averaging. We are particularly interested in the off–diagonal terms in $T_{ij}$ (of dimension $v^2$) because their gradients are responsible for nonconservation of the circulation. For comparison, these stresses are identically zero in inviscid fluids and have the form $\tau_{ij} = (\mu/\rho)\partial v_i/\partial x_j$ in moving viscous fluids. (To facilitate a direct comparison with $T_{ij}$, we have used the symbol $\tau_{ij}$, also of dimension $v^2$, for the stress tensor in fluids.) In stellar systems, the off–diagonal elements in $\partial T_{ij}/\partial x_j$ must be proportional to the velocity gradients for exactly the same reasons as in fluid systems (e.g. Landau and Lifshitz 1987).

Consider now the differences between the physical states of astrophysical fluid and stellar systems. The former are characterized by a short relaxation time towards local thermodynamical equilibrium (LTE) while the latter's relaxation time scale towards LTE is generally much longer than a Hubble time. This is due to the large difference in the frequency of collisions in these systems or, equivalently, because the mean free path between stellar collisions is much larger than the size of the system while in fluids it is much less. As a result, a fluid can be usually assumed to be in the LTE while a stellar system is only in the state of virial equilibrium after being subject to violent relaxation. The violent relaxation can leave the off–diagonal terms in $\partial T_{ij}/\partial x_j$ having the same order of magnitude as the diagonal terms, i.e. as the "pressure" gradients. [This is supported by observations of our Galaxy (e.g. Oort 1965; Kuijken and Gilmore 1989).] Such terms contribute to the acceleration in equation (4.2) and because of their dependence on velocity gradients they make a dominant contribution to $\nabla \times \mathbf{a}$ in equations (4.3) and (4.6).

The difference between rotating viscous fluids and stellar systems is also emphasized by a comparison between the magnitudes of "viscous" (off–diagonal) forces and of the inertial terms. Let us compare the Reynolds number of the flow in both cases. This number is defined as $Re \equiv vR/\nu$, where $v$ is the typical flow speed (e.g. the orbital speed in a disk), $R$ is the size of the system (i.e. the radius of a disk) and $\nu = \mu/\rho$ is the coefficient of kinematic viscosity for which $\nu \sim v_{rms}\ell$. Here $v_{rms}$ is the typical random velocity in the flow and $\ell$ is the mean free path between collisions. Hence, $Re \sim (v/v_{rms})(R/\ell) >> 1$ for a rotationally supported fluid disk and $Re \sim 1$ for an analogous stellar disk. This difference follows from the extremely different values of $\ell$ in the two types of systems.

The Reynolds number can also be written as the ratio of kinetic-to-"viscous" terms of the equation of motion. For stellar systems, we write not using the summation convention that

$$Re \sim \frac{\partial v_i v_j/\partial x_j}{\partial T_{ij}/\partial x_j} \sim 1. \tag{4.8}$$

This equation underlines the importance of "viscous" forces that are of the same order of magnitude as the conventional "pressure" gradients in shearing stellar systems. We conclude that the circulation is destroyed on time scales much longer than the dynamical time in viscous fluids (which leads to secular instability) and on time scales comparable to the dynamical time in stellar systems (which leads to dynamical instability). This dynamical instability is observed in $N$-body simulations of cold stellar disks but there also exists a well-known exception: stable rapidly rotating ($T/|W| > 0.14$) disks with a substantial fraction of particles on retrograde orbits.

If a fair fraction of particle orbits are reversed in a stellar–disk $N$–body model, then the total kinetic energy in ordered motion $T$ and and the total gravitational potential energy $W$ do not change but the total angular momentum $L$ decreases. We expect that the dynamical instability will be suppressed if the new value of $L$ falls below that at the bifurcation point of the stellar–disk Jacobi sequence (which occurs at $T/|W| \approx 0.14$ in the absence of reversed orbits). It is also clear from the discussions in §3.3 that if the mean rotation frequency $\Omega$ is arranged to be near zero, then the disk will evolve on a secular time scale toward the stellar–disk Dedekind sequence ($\Omega = 0$) because of gradual, slow mass loss (see also Figure 4 in Hunter 1974 where such evolutionary paths are shown).

The above physical interpretation of secular and dynamical instabilities, given in the context of phase transitions, illuminates the question of whether the bar instability of stellar systems is dynamical or secular. This question was discussed thoroughly by Vandervoort (1980, 1982, 1983; see also related work by Kalnajs 1972, Hunter 1979, Vandervoort & Welty 1982, and Vandervoort 1991) who concluded that the instability is dynamical. Our result substantiates the conclusion of Vandervoort and others. Furthermore, it supports a close dynamical similarity in the evolution of stellar and viscous fluid systems: circulation varies during the evolution of both types of systems but on different time scales [cf. equations (4.5) and (4.6)]. This similarity and its connection with the conservation law of circulation have not been previously noticed because of the apparent difference between the forms (4.1) and (4.2) of the equations of motion.

It is important to emphasize that the magnitudes of the gradients of "viscosity–like" terms in the stress tensor affect the time scale for development of the instability but not the point on the equilibrium sequence where the instability sets in. This point (the Jacobi bifurcation) is practically the same in viscous fluid and stellar systems ($T/|W| \approx 0.14$) because circulation is not conserved in both types of systems. On the other hand, the apparent difference between the Dedekind bifurcation in fluid ($T/|W| = 0.13753$) and in stellar ($T/|W| = 0$) systems has no physical significance because $T$ is defined differently in the two cases. In a Dedekind fluid with $\Omega = 0$, vortical motions are included in $T$ while, in a Dedekind stellar system, random azimuthal motions are excluded from $T$ [compare equations (2.5), (2.7) above to the corresponding equations in Freeman 1966].

### 4.3  Irrotational Motion and Superfluidity

So far in this paper, we have been concerned with second–order phase transitions on the Maclaurin sequence of oblate spheroids and their applications to astrophysical problems. On the other hand, the most notorious second–order phase transition in physics is the transition to a superconducting state and much of our experience about second–order phase transitions comes from studying superconductivity. The macroscopic theories of superconductivity and of superfluidity of liquid $^4$He were laid out by London (1950, 1954) in his monographs "Superfluids." In this subsection, we discuss briefly a common link between the above macroscopic theories and the second–order phase transitions on the Maclaurin sequence.

First, we note that magnetized plasmas are also constrained by a conservation law analogous to circulation conservation in fluids. The "frozen–in" condition is usually expressed in a form identical to that of equations (4.3) and (4.4), i.e.,

$$\frac{d\Phi_m}{dt} = \frac{d}{dt} \oint_c \mathbf{A} \cdot d\mathbf{c} = \frac{d}{dt} \int \int_S \mathbf{B} \cdot d\mathbf{S} = 0, \tag{4.9}$$

where $\mathbf{B} \equiv \nabla \times \mathbf{A}$ is the magnetic field, $\mathbf{A}$ is the magnetic vector potential, and $\Phi_m$ is the magnetic flux through surface $S$. When the conservation law (4.9) is violated the system makes a transition to turbulence. Like vorticity in fluids, the magnetic field is weakened by reconnection of small–scale loops and eddies and the magnetic energy of the system decreases. It seems plausible that in viscous fluids and in magnetized plasmas we may be witnessing the same nonlinear physical processes working on different time scales. The common link appears to be the violation on various time scales of the conservation law expressed through equations (4.4) and (4.9) above.

London (1950, 1954) realized that the macroscopic description of both superfluidity and superconductivity needed "new hydrodynamical equations." Specifically, he came to the conclusion that the zero–viscosity superfluid $^4$He could only be described if the classical hydrodynamical equations of motion were explicitly supplemented by the condition of irrotational motion (London 1954)

$$\nabla \times \mathbf{v} = 0. \tag{4.10}$$

Similarly, he formulated the equations of motion applicable to the superconducting state by imposing the "irrotational condition" (London 1950)

$$\nabla \times (\mathbf{v} + \frac{e}{mc}\mathbf{A}) = 0, \tag{4.11}$$

to the classical equations of motion of the nonviscous electronic fluid. In equation (4.11), $e$ and $m$ are the charge and the mass of the electron and $c$ is the speed of light. The term in parentheses is the effective "vorticity" of the electronic fluid. Equation (4.11) is simply a combination of equations (4.9) and (4.10).

In both cases that London analyzed, the phase transition leads to a minimum energy state free of "vorticity." Therefore, both phase transitions are similar to the *secular* instability on the Maclaurin sequence whose end–point is a Jacobi ellipsoid with no vortical motions. This does not mean that the $\lambda$–transition to superfluidity is of second order like the transition to superconductivity and the transition to the Jacobi sequence; it merely indicates that the final states (superconductor, superfluid, and Jacobi ellipsoid) share one common macroscopic property, i.e. they have zero "vorticity" (see also Paper II for a discussion of the $\lambda$–transition). On the other hand, the $\lambda$–transition does have some properties of a second–order phase transition. In that sense, we can understand the need for equations (4.10) and (4.11) in describing the macroscopic properties of the final equilibrium states of superconductivity and superfluidity.

The common link, expressed above through equations (4.4), (4.10), and (4.11), is not only limited to describing the "low–temperature" states of evolving "viscous" Maclaurin spheroids and superfluids. It also extends to the dynamics of the Jacobi ellipsoids themselves where the constraint (4.4) effects dramatic changes in the way evolution along the Jacobi sequence should be perceived. The results of our study concerning evolution along the Jacobi sequence will be presented in Paper II and in Paper III.

### 4.4  Spontaneous Symmetry Breaking and Phase Transitions

Bertin & Radicati (1976) were the first to point out that the bifurcation of nonaxisymmetric equilibrium sequences from the Maclaurin sequence constitutes a typical example of spontaneous symmetry breaking. This process has received a lot of attention in the past twenty years primarily as a result of its applications to gauge theories of fundamental interactions and to phase transitions in general. It turns out that self–gravitating, rotating equilibria have sufficiently rich structure to encompass many of the general features associated with spontaneous symmetry breaking and phase transitions (see also Paper II). More importantly, however, because the dynamics of rotating figures of equilibrium can be visualized and understood relatively easily, an examination of the mechanism of spontaneous symmetry breaking in this context may provide considerable insight about the mechanism itself.

The term "spontaneous symmetry breaking" is used to indicate the presence of solutions to a given dynamical problem with (spacetime or internal) symmetries lower than those of the Lagrangian (or free energy) from which the solutions were derived. Such solutions are generally invariant only to a subgroup of the original group of symmetries which leaves the Lagrangian of the problem invariant. Usually, this change in the symmetry of a solution is obtained as a parameter of the problem (e.g. the temperature or the energy) reaches a critical value. In gauge theories, this solution is called the "vacuum state" of a particular theory, while in the context of rotating fluids it is simply called a "figure of equilibrium." The breaking of the symmetry manifests itself with the appearance of an new observable, the order parameter, which vanishes in the higher–symmetry state and is also a measure of the departure from this state. In the gauge theories of fundamental interactions the order parameter is the "vacuum expectation value" of the so–called Higgs field whose role is precisely to break a given symmetry.

With these preliminaries, consider now rotating, self–gravitating figures of equilibrium. The free energy function of axisymmetric systems is symmetric with respect to rotations in two dimensions, i.e., under the SO(2) group of rotations about the symmetry axis (Bertin & Radicati 1976). The parameter that plays the role of "temperature" in this case is the inverse of the angular momentum $1/L$ (§4.1). Beyond a particular (critical) value of $L$, the minimum of the free energy is no longer obtained when the two equatorial axes are equal ($a = b$) but for $a \neq b$. Specifically, under conditions that conserve both angular momentum $L$ and circulation $C$, this critical temperature corresponds to the bifurcation point of the $x = +1$ self–adjoint Riemann sequence. Beyond this bifurcation point, the new equilibrium (i.e. the minimum energy state) is no longer axisymmetric and so the original SO(2) symmetry has been broken. The new equilibrium is now symmetric only under a discrete symmetry subgroup of the original group SO(2), i.e. it is symmetric only under rotations by $\pm 180°$ at every instant. The role of the order parameter is now played by the equatorial eccentricity $\eta = (1 - b^2/a^2)^{1/2}$ which is zero in the higher–symmetry state, i.e. on the Maclaurin sequence. Consider now the particular moment in time $t = t_o$ when symmetry breaks and the nonzero order parameter appears. The particular orientation of the ellipsoid in the inertial frame $X'Y'Z'$ can be expressed through the angle $\phi_o \equiv \phi(t_o)$ between the semimajor axis $a$ and the $X'$ axis. At any later time $t > t_o$, this orientation, $\phi \equiv \phi(t)$, is determined by the equation $\phi = \phi_o + \Omega t$, where $\Omega$ is the rotation frequency of the $x = +1$ Riemann ellipsoid. The angle $\phi_o$ itself is fixed arbitrarily at the instant of symmetry breaking $t = t_o$. The arbitrariness of the value of $\phi_o$, i.e. the arbitrariness of the orientation of the resulting ellipsoid at the instant $t = t_o$, is reminiscent of the massless (i.e. zero–frequency) Goldstone–boson modes of spontaneously broken symmetries of continuous groups (see Coleman 1988).

An analogous situation arises in cases where either $C$ or $L$ (but not both) is conserved. The absence of one conservation law (a "nonconserved current") allows now for the appearance of an energy minimum at a higher critical value of the "temperature" $1/L$, i.e. at a lower critical $L$. The resulting equilibria (Jacobi/Dedekind ellipsoids) have once again lower spatial symmetry since they are also nonaxisymmetric. The point of interest here is that the absence of a conserved current raises the critical temperature of the phase transition. It would be interesting to consider whether this behavior occurs in all second–order phase transitions and particularly in high–$T_c$ superconductivity.

Finally, the results obtained from the study of the stability on the Maclaurin sequence (presented above and in Paper II) support the following conclusion: a broken symmetry does not necessarily mean a change in *geometric symmetry*. Consider, for example, the three sequences of axisymmetric ring–like equilibria computed by Eriguchi & Sugimoto (1981; the one–ring sequence) and by Eriguchi & Hachisu (1983; the two–ring and the core–ring sequences). The stable ring–like configurations are all toroidal. Such toroidal equilibria do have lower energies than the corresponding Maclaurin spheroids but axisymmetry is not broken. Instead, the topology is broken in the following sense: The volume that a Maclaurin spheroid occupies in three–dimensional space is simply–connected while the volume of a toroidal configuration is doubly–connected if there is a hole in the middle (Lamb 1932). In fact, the connectedness of volume increases in the case of more than one ring structures. (We do not know whether field theories admit models with a similar property, i.e. models in which the different states differ in the topology/connectedness rather than in the symmetries of the associated group.) For toroidal/ring–like equilibria with "broken topology" and only one central hole, we may adopt the inner radius of the ring $R_{in}$ as the order parameter but the analogy with spontaneous symmetry breaking is only partially restored because of "degeneracy" between spheroids and tori: there exists one toroidal configuration that reaches the origin of the coordinate system and thus has an order parameter of $R_{in} = 0$ like any of the spheroids. This degeneracy is probably related to the fact that, in the framework of phase transitions, symmetry and topology breaking do remain quite different: the ring sequences of broken topology are related to a remarkable combination of a first–order and an apparent second–order phase transition unlike the Jacobi and the $x=+1$ sequences of broken axisymmetry that represent typical second–order phase transitions (see Paper II for details).

### 4.5  Conclusion, Possible Applications, and Future Work

Motivated by the works of Bertin & Radicati (1976) and Hachisu & Eriguchi (1983) who have pioneered the application of the theory of phase transitions to problems of stability in astrophysical fluids, we proceed to apply the same concept to all known bifurcation and instability points along equilibrium sequences of rotating, self–gravitating, systems. In this paper, we have presented results on second–harmonic instabilities and the related second–order, axisymmetry–breaking phase transitions in fluid and stellar systems. In the process, we have been able to clarify the nature of these instabilities and to examine the dynamical instability of stellar systems in relation to the nonconservation of circulation.

In future papers, we shall extend this study to phase transitions associated with axisymmetric and fourth–harmonic bifurcations (Paper II) and with third–harmonic bifurcations (paper III). As we mentioned above, we have already obtained surprising results concerning the fission hypothesis (e.g. Lebovitz 1972, 1987) and the breaking of topology and we anticipate more possible applications to the $\lambda$–transition of superfluid liquid $^4$He and to the standard classification scheme of thermodynamical phase transitions. We shall also address the question of an improved criterion for stability against second–harmonic disturbances in fluid and stellar systems.

### Acknowledgments

We thank G. Bertin, D. Brydges, G. Dandulakis, B. Deaver, Jr., M. Elitzur, W. Glatzel, G. Hess, P. Mannheim, and Q. Shafi for stimulating discussions and I. Hachisu, C. Hunter, N. Lebovitz, A. Toomre, and P. Vandervoort for useful correspondence. We are especially grateful to C. McKee and to R. Narayan for in–depth discussions of the physical concepts presented in this paper. IS is grateful to the Gauss Foundation for support and to K. Fricke, Director of Universitäts-Sternwarte Göttingen, for hospitality during a stay in which much of this work has been accomplished. IS thanks also the Center for Computational Studies of the University of Kentucky for continuing support. This work was supported in part by NASA grants NAGW–1510, NAGW–2447, NAGW–2376, and NAGW–3839, by NSF grant AST–9008166, and by grants from the San Diego Supercomputer Center and the National Center for Supercomputing Applications.

## APPENDIX

### Phase Transitions of Dynamical Systems

The secular and dynamical evolutionary paths that exist between equilibrium sequences of rotating, self–gravitating, incompressible fluids can be described not only through the properties of the free–energy function — as we have done in §3 of this paper — but also in terms of additional "thermodynamical" relations that exist between physical parameters such as the angular momentum $L$, the rotation frequency $\Omega$, and the specific moment of inertia $I$. "Thermodynamical meaning" can be assigned to such parameters by a straightforward comparison between the equations through which they are related and the corresponding equations of classical thermodynamics (e.g. Huang 1963). In what follows, we assume that changes take place at constant volume, a condition that implies mass conservation.

Identification of various quantities begins with the realization that nearly spherical systems with low angular momentum $L$ are stable and that instabilities (i.e. "phase transitions") appear as $L$ is increased along a particular equilibrium sequence. This increase in $L$ should be identified with a decrease in the temperature $\theta$ of a thermodynamical system, i.e.

$$\frac{1}{L} \to \theta, \tag{A1}$$

so that phase transitions appear in both cases when the temperature is decreased below a critical value. Two more analogies follow now immediately by dimensional comparison between the dynamical equation

$$L = I\Omega, \tag{A2}$$

and the thermodynamical equation of an ideal gas

$$U = c_{\mathrm{v}}\theta, \tag{A3}$$

where $U$ is the internal energy and $c_{\mathrm{v}}$ is the specific heat at constant volume. For an "ideal dynamical system" of constant $I$ analogous to an ideal gas of constant $c_{\mathrm{v}}$ we find that

$$I \to c_{\mathrm{v}}, \tag{A4}$$

and

$$\frac{1}{\Omega} \to U. \tag{A5}$$

Now we are in a position to calculate the "specific heat" $c_{\mathrm{v}}$ for a "nonideal" dynamical process using the thermodynamical definition $c_{\mathrm{v}} \equiv dU/d\theta$ and equations (A1), (A2), and (A5), i.e.

$$c_{\mathrm{v}} = \frac{d\Omega^{-1}}{dL^{-1}} = I^2 \frac{d\Omega}{dL}. \tag{A6}$$

Consider the behavior of the specific heat in the following cases: (a) In a smooth bifurcation, there is one tangent line to both sequences at the bifurcation point. So, both $d\Omega/dL$ and $I$ have the same values for both sequences at the bifurcation point and thus $c_{\mathrm{v}}$ is a continuous function of $L$. This indicates a "third–order phase transition." (b) In an abrupt bifurcation, $d\Omega/dL$ takes two different values for the two sequences at the bifurcation point and thus $c_{\mathrm{v}}$ will show a finite discontinuity at that point. This indicates a "second–order phase transition." (c) If a transition takes place "discontinuously" between sequences at constant $L$ (see Paper II), then equation (A6) shows that $c_{\mathrm{v}}$ goes to infinity at the transition point. This indicates either a "first–order phase transition" or a "$\lambda$–transition" (see Paper II).

The energy $E < 0$ is used in this paper as the "free–energy function." $E$ is equivalent to the thermodynamical Gibbs free energy $F$ for the following reason. A dimensional comparison between equations (A1), (A4), and the definition $c_{\mathrm{v}} \equiv d^2F/d\theta^2$ implies that

$$\frac{L^2}{I} \sim -E \to \frac{1}{F}. \tag{A7}$$

Thus, thermodynamics indicates that one should adopt the function $-1/E$ as the "free–energy function" of the dynamical system. However, the adopted free–energy function $E$ and its negative inverse $F$ are equivalent functions for our problem because they exhibit exactly the same monotonicity and the same extrema.

By analogy to the thermodynamical definition $s \equiv -dF/d\theta$, the "specific entropy" will generally be proportional to $dE/dL$, i.e.

$$s \sim -\frac{d(-E^{-1})}{dL^{-1}} \sim \frac{dE}{dL}. \qquad (A8)$$

The "specific entropy change" during a transition is calculated exactly from the thermodynamical equation $\theta ds = c_v d\theta$ and equations (A1), (A2), and (A6), i.e.

$$ds = -\frac{I^2}{L} d\Omega = -I d(\ln \Omega). \qquad (A9)$$

Finally, the "latent heat" $\mathcal{L}$ released during a "first–order phase transition" is found from the thermodynamical definition $\mathcal{L} \equiv \theta ds$, i.e.

$$\mathcal{L} = -\frac{I^2}{L^2} d\Omega = d\Omega^{-1}. \qquad (A10)$$

The "second–order phase transitions" discussed in this paper as well as "third–order phase transitions" (Paper II) appear at bifurcation points where $I, \Omega$ have the same values for both sequences. In such cases, equations (A9) and (A10) predict that $\Delta s = 0$ and $\mathcal{L} = 0$, respectively. Equivalently, we see from equation (A8) that the "specific entropy" is a continuous function in a "second–order phase transition" because $dE/dL$ takes the same value at the transition point (cf. Figure 4 above). Equation (A10) also suggests that "latent heat" is released only during a "phase transition" that takes place "discontinuously" between equilibrium sequences. Such a "discontinuous" evolution between sequences corresponds to either a "first–order phase transition" or to a "$\lambda$–transition" (see Paper II).

The role of circulation $C$ in the above analogy is not clear. We imagine that $C$ may be analogous to pressure $P$ in thermodynamics. In this case, phase transitions should appear with increasing "pressure" $C$ in our dynamical systems and the entire analogy may be formulated to also incorporate the analogy $C \rightarrow P$. However, this would not be useful since results are almost never presented or illustrated in terms of $C$. The "thermodynamical" significance of $C$ will also be discussed in Paper II in the context of "discontinuous phase transitions."

<div align="center">REFERENCES</div>